\documentclass[11pt]{article}
\usepackage[table]{xcolor}

\usepackage[preprint]{acl}

\usepackage{times}
\usepackage{latexsym}
\usepackage{multirow}
\usepackage{amssymb}
\usepackage{booktabs}
\usepackage{amsmath}
\usepackage{enumitem}
\usepackage{graphicx}

\usepackage{tcolorbox}
\tcbuselibrary{skins, breakable}

\newtcolorbox{promptbox}[2][]{
  enhanced,
  breakable,
  colback=white,
  colframe=black!70,
  colbacktitle=black!70,
  coltitle=white,
  title={#2},
  fonttitle=\bfseries,
  left=6pt, right=6pt, top=6pt, bottom=6pt,
  boxrule=0.6pt,
  arc=2mm,
  #1
}

\usepackage[T1]{fontenc}

\usepackage[utf8]{inputenc}

\usepackage{microtype}

\usepackage{inconsolata}

\usepackage{graphicx}
\usepackage{pifont}
\usepackage{tikz,xcolor}

\definecolor{SectionGray}{HTML}{ECEFF1}
\definecolor{LightBlue}{HTML}{F0F8FF}
\definecolor{lightyellow}{HTML}{FFF8BA}
\title{Uncertainty-based Debiasing and Unlearning for Decontamination}

\author{
\textbf{Guangzhi Sun\textsuperscript{1}},
 \textbf{Xiao Zhan\textsuperscript{1,2}},
 \textbf{Mark Gales\textsuperscript{1}}
\\
\\
 \textsuperscript{1}Department of Engineering, University of Cambridge \\
 \textsuperscript{2}VRAIN, Universitat Politècnica de València 
\\
 \texttt{gs534@cam.ac.uk}
}

\begin{document}
\maketitle
\begin{abstract}

Benchmark-based evaluation is the dominant paradigm for assessing large language model (LLM) capabilities, yet data contamination inflates reported performance and undermines fair comparison. Existing decontamination methods are evaluated solely through aggregate accuracy, which can obscure substantial differences in per-sample model behaviour, and many require access to an uncontaminated model. In this paper, we propose a sample-level evaluation framework for decontamination that complements accuracy-based assessment with distributional distance metrics, measuring how closely a decontaminated model recovers the output distribution of an uncontaminated model on each sample. Building on this framework, we introduce Uncertainty-Based Decontamination (UBD), a family of methods that leverage deep ensembles of the contaminated model to estimate per-sample memorization without requiring a uncontaminated model or knowledge of which samples are contaminated. UBD estimates a per-sample correction scalar from ensemble uncertainty, which is used to construct a debiased target distribution that suppresses the inflated probability mass on correct answers induced by contamination. This target is then used either as a post-hoc output correction (debiasing) or as a soft training signal for parameter update (unlearning). Experiments on MMLU-Pro and MATH-MCQA across multiple LLM backbones demonstrate that UBD produces per-sample output distributions substantially closer to those of an uncontaminated model than paraphrasing or choice-permutation baselines, while preserving model performance on uncontaminated data.\footnote{\url{https://github.com/BriansIDP/UnlearnForDataContamination.git}}

\end{abstract}

\section{Introduction}
Large language models (LLMs) \cite{llama,mistral,qwen25,qwen3,gemini25} have attained a level of capability that enables them to perform a broad range of tasks across diverse domains. As these models continue to be developed, released, and deployed at scale, the need for a systematic framework to evaluate and compare their capabilities has become increasingly important. In this context, benchmark-based evaluation has emerged as a standard paradigm for assessing LLM performance. Benchmarks such as MMLU \cite{mmlu,mmlupro} evaluate models on a collection of samples and report performance by comparing model outputs with reference answers. However, data contamination \cite{deng2024investigating,datacontamination2} presents a particularly significant challenge for LLMs due to the vast scale of their training data, which makes it difficult to ascertain whether specific benchmark samples have been used for training. As a result, data contamination can inflate reported performance and lead to unfair comparisons, thereby obscuring the true capabilities of the models under evaluation.


To ensure fair and trustworthy benchmarking of large language models, recent work has explored a range of decontamination methods. Early approaches mitigate the effects of contaminated data through sampling strategies \cite{dong2024generalization} or paraphrasing-based techniques \cite{itd}, both of which aim to reduce reliance on verbatim memorization of reference answers. Subsequent research shifted toward modifying the contaminated model itself, either by identifying and removing neural shortcuts \cite{neuronshort} or by directly minimizing the KL divergence from an uncontaminated uncontaminated model \cite{chai2026benchmarks}. Despite this progress, the evaluation of decontamination methods largely centers on dataset-level performance metrics, such as accuracy, which can obscure important changes in behavior at the level of individual samples. For instance, two models may achieve identical accuracy on the same dataset while producing correct predictions on entirely disjoint subsets of examples. This limitation is especially concerning in settings where the specific contaminated portion of the data is unknown.

Therefore, we propose to evaluate decontamination efficacy at the sample level that complements accuracy-based assessment with distribution-distance measures. We focus primarily on multiple-choice question (MCQ) benchmarks, one of the most widely used evaluation formats for LLMs. The proposed evaluation framework captures the behavioral similarity between decontaminated and uncontaminated models on each sample. Building on this, we introduce uncertainty-based decontamination (UBD) approaches via de-biasing and unlearning. These approaches do not require access to a uncontaminated model and do not require knowledge of the pre-training dataset. By estimating uncertainty through deep ensembles, UBD produces per-sample output distributions that are substantially closer to those of an uncontaminated model than those obtained with paraphrasing or choice-permutation baselines. Furthermore, uncertainty provides a useful signal for distinguishing hard but memorized samples from genuinely easy ones, even when both exhibit low loss after contamination. Contributions are summarized as follows.

\vspace{-0.2cm}
\begin{itemize}[leftmargin=*]
    \item We argue that dataset-level metrics can appear close while masking substantial differences in how the model actually behaves on individual samples. Rather than relying only on dataset-level performance, we propose measuring decontamination effectiveness at the per-sample level using distribution-distance metrics.
    \vspace{-0.2cm}
    \item We propose uncertainty metrics to measure the degree of contamination. These metrics are important to distinguish between hard but memorized samples and genuinely easy samples
    \vspace{-0.2cm}
    \item We propose UBD approaches via de-biasing and unlearning as effective LLM decontamination methods that leverage uncertainty estimation from deep ensembles, without requiring an uncontaminated or uncontaminated model.
\end{itemize}

\section{Related Work}

\subsection{Data Contamination and Detection}
Data contamination occurs when benchmark samples are included in an LLM’s training corpus, leading to inflated evaluation results that do not accurately reflect the model’s true capabilities~\cite{balloccu2024leak,deng2024investigating,chang2024survey,sainz2023nlp}. For instance, in the context of mathematical reasoning, \cite{xu2024benchmarking} analyzed 31 LLMs and found evidence of widespread contamination. Prior work has also further distinguished two forms of benchmark contamination: exact contamination, in which test instances appear verbatim in the training data, and syntactic contamination, in which test instances reappear in superficially altered forms, such as paraphrases or prompt-prefixed variants~\cite{datacontamination2}.


Existing contamination detection methods can be broadly grouped into two paradigms. One line of work relies on model-output statistics, using signals such as peaked output distributions~\cite{dong2024generalization} or anomalous token probabilities~\cite{shi2023detecting} to infer whether a test instance has likely appeared in training. However, recent evidence~\cite{fu2025does} suggests that such membership-style signals may become much less informative at pretraining scale. A second line of work instead probes generalization behavior, asking whether the model can reconstruct or recover withheld answer information under controlled perturbations~\cite{deng2024investigating}. This perspective has also revealed harder-to-detect forms of contamination, including cross-lingual variants that are not easily captured by surface-level matching~\cite{yao2024data}.

\subsection{Mitigation for Contamination}

Contamination mitigation, or decontamination, has been mainly explored via dynamic benchmarking or model-level modification. The former prevents contamination by constructing evaluation samples guaranteed to be unseen \citep{wu2025antileakbench} or by rewriting leaked samples via paraphrasing at inference time \citep{itd}. The latter corrects deployed models either by replacing shortcut neurons \citep{neuronshort} or by minimising KL divergence toward a clean reference \citep{chai2026benchmarks,yao2024machine, lizzo2025unlearn}, both of which require an uncontaminated uncontaminated model and evaluate only at the benchmark-level accuracy. On the contrary, our approach does not require access to a uncontaminated model or prior knowledge of the contaminated portion of data, and we evaluate at the contamination sample level via per-sample distribution distances.

\section{Problem Formulation}
\label{sec:prob}

Let $f(\cdot\,; \theta): \mathcal{X} \rightarrow \mathcal{Y}$ denote a language model with parameters $\theta$, mapping an input $x \in \mathcal{X}$ to a probability distribution over the output space $\mathcal{Y}$. For a given input $x$, we write $P(y \mid x; \theta)$ for the probability assigned to output sequence $y \in \mathcal{Y}$.

Let $\mathcal{D}_{\text{test}} = \{(x^{(i)}, y^{(i)}_*)\}_{i=1}^N$ be a benchmark to test, where $x^{(i)}$ is the $i$-th test input and $y^{(i)}_*$ is its ground-truth response. We denote the model's output distribution as $p^{(i)}(\cdot) = P(\cdot \mid x^{(i)}; \theta)$.

\subsection{Data Contamination}

A model with \textit{uncontaminated} parameters $\theta$ is one trained on a corpus $\mathcal{D}_\text{train}$ that does not include $\mathcal{D}_{\text{test}}$. A \textit{contamination} process introduces some part of $\mathcal{D}_{\text{test}}$ into training, yielding \textit{contaminated} parameters $\bar{\theta}$. We denote the part of $\mathcal{D}_\text{test}$ that is used for training as $\mathcal{D}_\text{eval}$, and the rest as $\mathcal{D}_\text{dev}$, the contamination process is thus:
\begin{equation}
    \theta \;\xrightarrow{\;\text{train on } \mathcal{D}_{\text{train}}\cup \mathcal{D}_{\text{eval}}\;}\; \bar{\theta}
\end{equation}
We denote the output distributions of the uncontaminated and contaminated models on $x^{(i)}$ as $p^{(i)}(\cdot) = P(\cdot \mid x^{(i)}; \theta)$ and $\bar{p}^{(i)}(\cdot) = P(\cdot \mid x^{(i)}; \bar{\theta})$ respectively.
Data contamination causes $\bar{p}^{(i)}$ to concentrate probability mass on $y^{(i)}_*$, inflating performance on $\mathcal{D}_{\text{eval}}$ and hence $\mathcal{D}_\text{test}$ beyond what reflects genuine model capability. For instance, in multiple-choice questions, data contamination manifests as inflated probabilities for the correct choices. In open-ended generation, it manifests as near-verbatim reproduction of reference outputs.

\subsection{Decontamination}

We define \textit{decontamination} as the task of recovering, from the contaminated model $\bar{\theta}$, an output distribution that approximates that of the unbiased model for all inputs in $\mathcal{D}_{\text{test}}$. Formally, we seek a corrected distribution $\hat{p}^{(i)}(\cdot)$ such that:
\begin{equation}
    \hat{p}^{(i)}(\cdot) \approx p^{(i)}(\cdot) \quad \forall\, x^{(i)} \in \mathcal{D}_{\text{test}}
\end{equation}
We constrain the decontamination procedure to have access only to $\bar{\theta}$ (or a deep ensemble of $\bar{\theta}$) and $\mathcal{D}_{\text{test}}$. Neither the unbiased parameters $\theta$ nor the clean distributions $p^{(i)}$ are assumed to be available. We further constrain that we do not know which part of $\mathcal{D}_\text{test}$ is contaminated, and hence decontamination should be applied to all the samples in $\mathcal{D}_\text{test}$. These constraints reflect a practical scenario, posing more challenges to the methodologies.

We consider debiasing and unlearning as two types of decontamination methods. \textit{Output debiasing} directly corrects $\bar{p}^{(i)}$ at inference time without modifying the model parameters, producing a calibrated distribution $\hat{p}^{(i)}$ as a function of $\bar{p}^{(i)}$:
\begin{equation}
    \hat{p}^{(i)}(\cdot) = g\!\left(\bar{p}^{(i)}(\cdot)\right)
    \label{eq:debiasing}
\end{equation}
where $g(\cdot)$ is a correction function applied post-hoc that either redistributes the output distribution $\bar{p}^{(i)}$ directly or modifies input $x^{(i)}$ to influence $\bar{p}^{(i)}$, such as paraphrasing-based methods.

\textit{Unlearning} instead updates the model parameters via a targeted training objective to suppress memorization of $\mathcal{D}_{\text{eval}}$, yielding purified parameters $\hat{\theta}$ and correspondingly $\hat{p}^{(i)}(\cdot) = P(\cdot \mid x^{(i)}; \hat{\theta})$:
\begin{equation}
    \bar{\theta} \xrightarrow{\text{unlearn}} \hat{\theta}
    \label{eq:unlearnconcept}
\end{equation}
Both families share the same goal of recovering $p^{(i)}$ and are evaluated under the same metrics.

\subsection{Measuring Decontamination Effectiveness}

Prior work evaluates decontamination by measuring the reduction in aggregate accuracy on $\mathcal{D}_{\text{test}}$ \cite{itd,dong2024generalization,neuronshort,chai2026benchmarks}. \textit{We argue this is insufficient}: A method may yield similar overall accuracy by being correct on two completely different subsets of data. We instead propose evaluating decontamination at the \textit{per-sample level}.

Let $\hat{p}^{(i)}(\cdot) = P(\cdot \mid x^{(i)}; \hat{\theta})$ denote the output distribution of the decontaminated model. We measure effectiveness via the following metrics.

\paragraph{Per-Sample Distribution Distances.} We use the KL-divergence to compare the full output distributions between the uncontaminated and decontaminated models at the per-sample level:
\begin{align}
    D_{\text{KL}} &= \frac{1}{N}\sum_{i=1}^{N} D_{\text{KL}}\!\left(p^{(i)} \,\|\, \hat{p}^{(i)}\right)
\end{align}

\paragraph{Probability Difference of Ground-truth.} We further measure how closely the decontaminated model recovers the unbiased probability assigned to the ground-truth response $y^{(i)}_*$:
\begin{equation}
    D_\text{L1}^* = \frac{1}{N}\sum_{i=1}^{N} \left|\hat{p}^{(i)}(y^{(i)}_*) - p^{(i)}(y^{(i)}_*)\right|
\end{equation}
By focusing on the ground truth, we can directly assess the degree of probability inflation that remains after decontamination without introducing noise. Specifically, in multiple-choice tasks, $y^{(i)}_*$ corresponds to the letter of the correct answer, making $p^{(i)}(y^{(i)}_*)$ directly interpretable as the model's confidence in the correct choice. A decontamination method is considered effective if it minimizes both metrics simultaneously, recovering the unbiased distribution at the sample level rather than merely closing the aggregate accuracy gap. 

\paragraph{Preserving performance on other data.} A decontamination method that recovers $p^{(i)}$ on $\mathcal{D}_{\text{eval}}$ but degrades model performance on $\mathcal{D}_\text{dev}$ is of limited practical value. We therefore additionally evaluate the decontaminated model on $\mathcal{D}_{\text{dev}}$ with the same distributional metrics $D_\text{KL}$ and $D_\text{L1}^*$ to measure the deviation of $\hat{p}^{(i)}$ from $\bar{p}^{(i)}$ on $\mathcal{D}_{\text{dev}}$. Note that here we use $\bar{p}^{(i)}$ specifically to measure the deviation from the deployed model checkpoint.

\section{Uncertainty-Based Decontamination}

\begin{figure*}[t]
    \centering
    \includegraphics[width=0.9\linewidth]{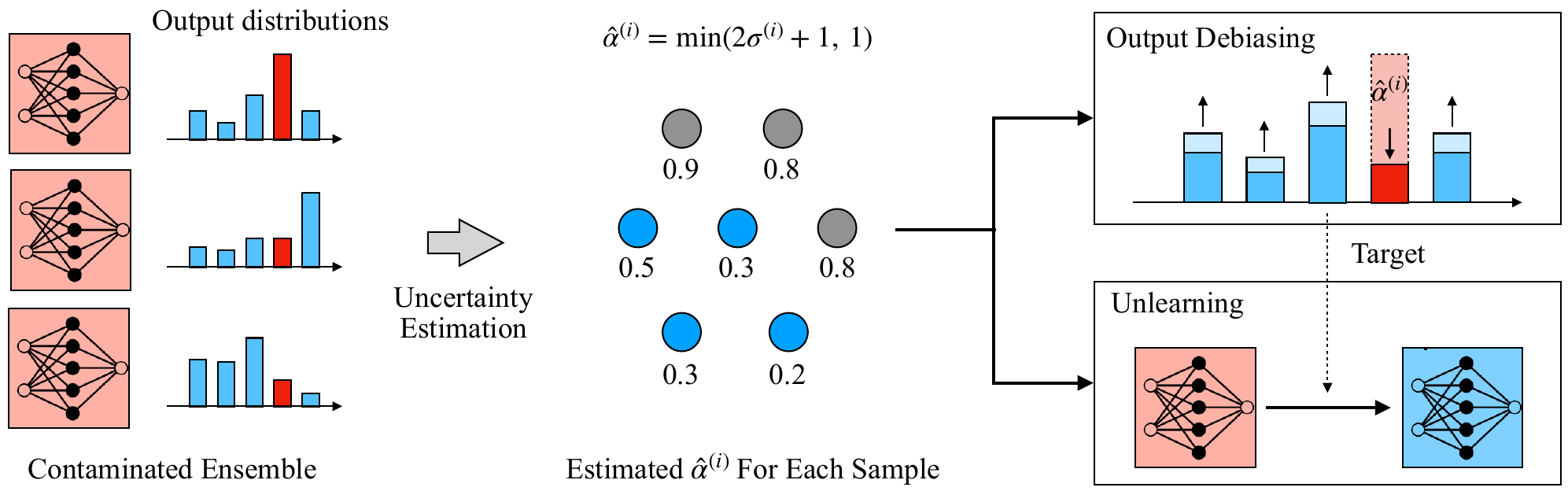}
    \caption{Illustration of the uncertainty-based decontamination (UBD) approach. Uncertainty is first estimated from the ensemble, and the level of contamination $\hat{\alpha}^{(i)}$ is then estimated using the uncertainty. Debiasing is applied by suppressing the ground-truth probability and increasing other probabilities proportionally, and unlearning can then be applied using the debiased output as the target.}
    \vspace{-0.3cm}
    \label{fig:ubd}
\end{figure*}

We now introduce UBD approaches which are directly applied to each sample in $\mathcal{D}_\text{test}$ under constraints defined in Section \ref{sec:prob}. The process is illustrated in Fig. \ref{fig:ubd}. First, output distributions are obtained from the ensemble of contaminated models, and knowledge uncertainty is computed for each sample. An estimation of the degree of contamination is obtained from uncertainty (denoted as $\alpha$ introduced in Section \ref{sec:debiasing}). Then, debiasing can be applied to rescale the probabilities according to the degree of contamination. Alternatively, unlearning can be performed to change the model parameters using the rescaled distributions as targets. 

\subsection{UBD-Debiasing}
\label{sec:debiasing}

UBD-Debiasing corrects the contaminated distribution $\bar{p}^{(i)}$ at inference time without modifying model parameters. The core of debiasing is to suppress the probability mass on the ground-truth response $y^{(i)}_*$, which has been inflated during contamination, while equally increasing the probability on other choices while keeping the relative shape of the distribution over remaining options unchanged. We achieve this by introducing a per-sample scalar $\alpha^{(i)}$ that captures the degree of contamination, defined as the ratio of the uncontaminated to the contaminated ground-truth probability in Eqn \eqref{eq:alpha}.

\begin{equation}
    \alpha^{(i)} = \frac{p^{(i)}(y^{(i)}_*)}{\bar{p}^{(i)}(y^{(i)}_*)}
    \label{eq:alpha}
\end{equation}

Since the uncontaminated ground-truth probability is unknown, we estimate $\hat{\alpha}^{(i)}$ using uncertainty in Section \ref{sec:unc}. We then construct a debiased distribution $\hat{p}^{(i)}$ by rescaling the ground-truth probability and redistributing the remaining mass proportionally across the other outputs.

\begin{equation}
\small
    \hat{p}^{(i)}(y) = \begin{cases}
        \hat{\alpha}^{(i)} \,\bar{p}^{(i)}(y^{(i)}_*) \quad \quad \text{if } y = y^{(i)}_* \\
        \dfrac{1 - \hat{\alpha}^{(i)}\,\bar{p}^{(i)}(y^{(i)}_*)}{1 - \bar{p}^{(i)}(y^{(i)}_*)} \cdot \bar{p}^{(i)}(y) \quad\text{o.w.}
    \end{cases}
\end{equation}

This correction function $g(\bar{p}^{(i)})$ (cf.\ Eqn~\eqref{eq:debiasing}) is applied post-hoc and requires no access to an uncontaminated model or pre-training data.

\subsection{UBD-Unlearning}

UBD-Unlearning updates the contaminated model parameters $\bar{\theta}$ via a targeted training objective, using the debiased distribution $\hat{p}^{(i)}$ constructed above as the training target. Concretely, we fine-tune $\bar{\theta}$ by minimising the cross-entropy between the model's current output and the debiased target distribution over all samples in $\mathcal{D}_\text{test}$ as shown in Eqn. \eqref{eq:unlearn}.
\begin{equation}
    \mathcal{L} = -\frac{1}{N}\sum_{i=1}^{N} \sum_{y} \hat{p}^{(i)}(y) \log P(y \mid x^{(i)}; \hat{\theta})
    \label{eq:unlearn}
\end{equation}
This drives the model toward the debiased output distribution $\hat{p}^{(i)}$ rather than toward hard labels, allowing gradient updates to suppress the inflated probability on $y^{(i)}_*$ while preserving the overall shape of the output distribution. The resulting decontaminated parameters $\hat{\theta}$ (cf.\ Eqn~\eqref{eq:unlearnconcept}) are directly used for inference. Since decontamination should be applied to all samples in $\mathcal{D}_\text{test}$, including those that may not be contaminated, the debiased target $\hat{p}^{(i)}$ defaults to $\bar{p}^{(i)}$ when $\hat{\alpha}^{(i)} \approx 1$, making the training signal negligible for clean samples. Both debiasing and unlearning are evaluated under the $D_\text{KL}$ and $D_\text{L1}^*$ metrics on both $\mathcal{D}_\text{eval}$ and $\mathcal{D}_\text{dev}$.

\subsection{Estimating $\alpha$ from Uncertainty}
\label{sec:unc}


In a contaminated model, high performance can result from two different scenarios. The first is from non-contaminated training data that enables the model to accurately predict the answer (i.e. easy samples). The second scenario is that there is a contaminated sample that results in significantly higher performance than the uncontaminated model (i.e. hard but contaminated samples).
In the first case, where we have multiple non-contaminated training samples to support, there is little sensitivity to the precise batch ordering of the relevant training samples. Conversely, as memorization is highly related to the \textit{batch ordering} \cite{lesci-etal-2024-causal}, there is expected to be more variability in the predictions of the ensemble members. This then results in the unusual balance of high confidence prediction and high variability among ensemble members, which is related to the epistemic uncertainty.

We assume access to an ensemble of $M$ contaminated models $\{\bar{\theta}_1, \ldots, \bar{\theta}_M\}$, obtained by training with identical hyperparameters but different batch orderings. We propose two complementary uncertainty signals derived from this ensemble to estimate the per-sample contamination scalar $\hat{\alpha}^{(i)}$.

Due to the variable bound of the knowledge uncertainty as defined in Appendx \ref{sec:uncertainty}, (i.e. $[0, \log{|\mathcal{Y}|}]$), we employ \textbf{standard deviation of the ground-truth probability} $\bar{p}^{(i)}(y^{(i)}_*)$ across ensemble members as a simpler uncertainty signal:
\begin{align}
    \mu^{(i)} & = \frac{1}{M}\sum_{m=1}^{M}\bar{p}_{m}^{(i)}(y^{(i)}_*), \\
    \sigma^{(i)} &= \sqrt{\frac{1}{M}\sum_{m=1}^{M} 
    \left(\bar{p}_m^{(i)}(y^{(i)}_*) - \mu^{(i)}\right)^{2}}
    \label{eq:sigma}
\end{align}
We observe a strong negative correlation between $\sigma^{(i)}$ and the oracle $\alpha^{(i)}$: samples with higher standard deviation in the ground-truth probability across the ensemble tend to have lower $\alpha^{(i)}$, indicating stronger memorization. This is because memorized samples are more sensitive to the order in which contaminated data is presented in training.

\paragraph{Estimating $\hat{\alpha}^{(i)}$.} We exploit the above correlation to derive a parameter-free estimate of $\alpha^{(i)}$ from ensemble standard deviation. Specifically, we set:
\begin{equation}
    \hat{\alpha}^{(i)} = \begin{cases}
        1 - 2\sigma^{(i)} \quad \quad \text{if } \bar{p}^{(i)}(y_*^{(i)}) \geq T_p \\
        1\quad\text{otherwise}
    \end{cases}
    \label{eq:alpha_pred}
\end{equation}
The coefficient of 2 bounds the prediction between 0 and 1 since $\sigma^{(i)}\in [0, 0.5)$, and $T_p\in[0, 1)$ is the threshold above which decontamination is applied. Applying $T_p$ excludes samples where the contaminated model is already performing poorly. That is, for a particular member of the ensemble, we penalize the high-confidence correct predictions from that model. This estimator requires no ground-truth $\alpha$ values for calibration and applies to any sample in $\mathcal{D}_\text{test}$, including those that are not contaminated, for which $\sigma^{(i)} \approx 0$ and hence $\hat{\alpha}^{(i)} \approx 1$, leaving the output distribution unchanged.

\begin{table*}[t]
    \centering
        \caption{Results on {Llama-3.2-3B-Instruct} and Qwen2.5-3B-Instruct on \textbf{MMLU-pro}. The target values are highlighted in yellow cells, and the closest results to the target are in bold. The threshold $T_p$ is defined in Eqn \eqref{eq:alpha_pred} above which decontamination is applied. $\mathcal{D}_\text{eval}$ metrics are computed against the uncontaminated model, while $\mathcal{D}_\text{dev}$ metrics are computed against the model before decontamination. RC is the residual contamination, which is the absolute difference between the uncontaminated model and the decontaminated model.}
    \resizebox{0.92\textwidth}{!}{%
    \begin{tabular}{p{8cm}cccccc}
    \toprule
    \multirow{2}{*}{Method}     & \multicolumn{2}{c}{Accuracy (RC $\downarrow$)} &  \multicolumn{2}{c}{$D_\text{KL}~\downarrow$} & \multicolumn{2}{c}{$D_\text{L1}^*~\downarrow$} \\
    & $\mathcal{D}_\text{dev}$ & $\mathcal{D}_\text{eval}$ & $\mathcal{D}_\text{dev}$ & $\mathcal{D}_\text{eval}$ & $\mathcal{D}_\text{dev}$ & $\mathcal{D}_\text{eval}$ \\
    \midrule
    \rowcolor{SectionGray} Llama-3.2-3B-Instruct & & & & & & \\
    Uncontaminated model & -- & \cellcolor{lightyellow} 43.4 (0.0) & -- & \cellcolor{lightyellow} 0 & -- & \cellcolor{lightyellow} 0 \\
    Contaminated Model & \cellcolor{lightyellow} 42.2 (0.0) & 60.6 (17.2) & \cellcolor{lightyellow} 0 & 0.8411 & \cellcolor{lightyellow} 0 &	0.2311 \\
    Choice Permutation & 41.4 (0.8) & 53.1 (9.7) & 0.5658 & 0.9515 & 0.1486 & 0.2307 \\
    Paraphrasing \cite{itd} & \textbf{42.3} (\textbf{0.1}) & 58.1 (14.7) & 0.1328 & 0.8297 & \textbf{0.0637} & 0.2252\\
    Choice Permutation + Paraphrasing & 41.5 (0.7) & 51.8 (8.4) & 0.6415 & 0.9580 & 0.1666 & 0.2296 \\
    DeconIEP \cite{chai2026benchmarks} & 32.0 (10.2) & 42.7 (0.7) & 0.2909 & 0.6201 & 0.1920 & 0.2187 \\
    \rowcolor{LightBlue} UBD-Debiasing ($T_p = 0.0$)  & 30.0 (12.2) & \textbf{43.7} (\textbf{0.3}) &	0.0894 &	0.5597 &	0.1063 &	\textbf{0.1482} \\
    \rowcolor{LightBlue} UBD-Unlearning ($T_p = 0.0$) & 33.3 (8.9) & 52.1 (8.7) & 0.1689 & \textbf{0.5184} & 0.1196 & 0.1656 \\
    \rowcolor{LightBlue} UBD-Debiasing ($T_p = 0.5$) & 36.4 (5.8) & 51.7 (8.3) & \textbf{0.0749} & 0.5253 & {0.0753} & 0.1486 \\
    \rowcolor{LightBlue} UBD-Unlearning  ($T_p = 0.5$) & 33.1 (9.1) & 52.0 (8.6) & 0.1710 & 0.5192 & 0.1218 & 0.1651\\
    \midrule
    \rowcolor{SectionGray} Qwen2.5-3B-Instruct & & & & & & \\
        Uncontaminated model & -- & \cellcolor{lightyellow} 44.8 (0.0) & -- & \cellcolor{lightyellow} 0 & -- & \cellcolor{lightyellow} 0 \\
    Contaminated Model & \cellcolor{lightyellow} 46.0 (0.0) & 69.4 (24.6) & \cellcolor{lightyellow} 0 & 1.3777 & \cellcolor{lightyellow} 0 & 0.2867 \\
    Choice Permutation & \textbf{45.1} (\textbf{0.9}) & 55.1 (10.3) & 1.2480 & 1.5128 & 0.1906 & 0.2610 \\
    Paraphrasing \cite{itd} & 45.1 (0.9) & 64.3 (19.5) & 0.3567 & 1.3846 & \textbf{0.0990} & 0.2779 \\
    Choice Permutation + Paraphrasing & 44.3 (1.7) & 53.3 (8.5) & 1.0785 & 1.5579 & 0.1213 & 0.2655 \\
    DeconIEP \cite{chai2026benchmarks} & 35.6 (10.4) & \textbf{49.8} (\textbf{5.0}) & 0.3245 & 0.8166 & 0.1725 & 0.2503 \\
    \rowcolor{LightBlue} UBD-Debiasing ($T_p = 0.0$)  & 33.3 (12.7) & {51.2} ({6.4}) & 0.1436 & 0.6518 & 0.1225 & \textbf{0.1498} \\
    \rowcolor{LightBlue} UBD-Unlearning ($T_p = 0.0$) & 38.0 (8.0) & 57.5 (12.7) & 0.3381 & \textbf{0.6147} & 0.1513 & 0.1896 \\
    \rowcolor{LightBlue} UBD-Debiasing ($T_p = 0.5$) & 36.6 (9.4) & 54.9 (10.1) & \textbf{0.1331} & 0.6235 & {0.1028} & 0.1499 \\
    \rowcolor{LightBlue} UBD-Unlearning  ($T_p = 0.5$) & 38.4 (7.6) & 57.2 (12.4) & 0.3337 & 0.6370 & 0.1463 & 0.1909 \\
    \bottomrule
    \end{tabular}
    }
    \vspace{-0.3cm}
    \label{tab:llamammlupro}
\end{table*}

\begin{table*}[t]
    \centering
        \caption{Decontamination results using {Llama-3.2-3B-Instruct} on \textbf{MATH-MCQA}. The target values are highlighted in yellow cells, and the closest results to the target are in bold. The threshold $T_p$ is defined in Eqn \eqref{eq:alpha_pred} above which decontamination is applied. $\mathcal{D}_\text{eval}$ metrics are computed against the uncontaminated model, while $\mathcal{D}_\text{dev}$ metrics are computed against the contaminated model before decontamination. RC is the residual contamination, which is the absolute difference between the uncontaminated model and the decontaminated model.}
    \resizebox{0.92\textwidth}{!}{%
    \begin{tabular}{p{8cm}cccccc}
    \toprule
    \multirow{2}{*}{Method}     & \multicolumn{2}{c}{Accuracy (RC)} &  \multicolumn{2}{c}{$D_\text{KL}$} & \multicolumn{2}{c}{$D_\text{L1}^*$} \\
    & $\mathcal{D}_\text{dev}$ & $\mathcal{D}_\text{eval}$ & $\mathcal{D}_\text{dev}$ & $\mathcal{D}_\text{eval}$ & $\mathcal{D}_\text{dev}$ & $\mathcal{D}_\text{eval}$ \\
    \midrule
    \rowcolor{SectionGray} Llama-3.2-3B-Instruct & & & & & & \\
    Uncontaminated model & -- & \cellcolor{lightyellow} 55.3 (0.0) & -- & \cellcolor{lightyellow} 0 & -- & \cellcolor{lightyellow} 0 \\
    Contaminated Model & \cellcolor{lightyellow} 54.6 (0.0) & 77.4 (22.1) & \cellcolor{lightyellow} 0 & 1.0305 & \cellcolor{lightyellow} 0 & 0.2660 \\
    Choice Permutation & 53.9 (0.7) & 60.8 (5.5) & 1.1781 & 1.3272 & 0.2666 & 0.3147 \\
    Paraphrasing \cite{itd} & 54.1 (0.5) & 72.2 (16.9) & 0.2027 & 0.9088 & \textbf{0.1075} & 0.2551 \\
    Choice Permutation + Paraphrasing & \textbf{54.4} (\textbf{0.2}) & 58.5 (3.2) & 1.2465 & 1.2580 & 0.2976 & 0.3084 \\
    DeconIEP \cite{chai2026benchmarks} & 43.6 (11.0) & 59.2 (3.9) & 0.1837 & 0.4049 & 0.1541 & 0.2162 \\
    \rowcolor{LightBlue} UBD-Debiasing ($T_p = 0.0$) & 36.1 (18.5) & \textbf{56.8} (\textbf{1.5}) & 0.1614 & 0.3687 & 0.1577 & 0.1639 \\
    \rowcolor{LightBlue} UBD-Unlearning ($T_p = 0.0$) & 47.8 (6.8) & 65.3 (10.0) & 0.3114 & 0.3653 & 0.1978 & 0.1947 \\
    \rowcolor{LightBlue} UBD-Debiasing ($T_p = 0.5$) & 39.1 (15.4) & 59.9 (4.6) & \textbf{0.1467} & \textbf{0.3332} & 0.1291 & \textbf{0.1564} \\
    \rowcolor{LightBlue} UBD-Unlearning  ($T_p = 0.5$) &  48.5 (6.1) & 65.3 (10.0) & 0.3062 & 0.3404 & 0.2114 & 0.1955 \\
    \rowcolor{SectionGray} Qwen2.5-3B-Instruct & & & & & & \\

    Uncontaminated model & -- & \cellcolor{lightyellow} 65.8 (0.0) & -- & \cellcolor{lightyellow} 0 & -- & \cellcolor{lightyellow} 0 \\
    Contaminated Model & \cellcolor{lightyellow} 65.4 (0.0) & 85.6 (19.8) & \cellcolor{lightyellow} 0 & 1.0010 & \cellcolor{lightyellow} 0 & 0.2685 \\
    Choice Permutation & 64.5 (0.9) & 71.6 (5.8) & 0.9244 & 1.2529 & 0.2419 & 0.3045 \\
    Paraphrasing \cite{itd} & 65.7 (0.3) & 81.2 (15.4) & 0.1743 & 0.9408 & \textbf{0.1065} & 0.2627 \\
    Choice Permutation + Paraphrasing & \textbf{65.2} (\textbf{0.2}) & 70.0 (4.2) & 1.0093 & 1.2072 & 0.2591 & 0.2974 \\
    DeconIEP \cite{chai2026benchmarks} & 55.6 (9.8) & 70.7 (4.9) & 0.2576 & 0.6153 & 0.1632 & 0.2589 \\
    \rowcolor{LightBlue} UBD-Debiasing ($T_p = 0.0$) & 50.1 (15.3) & \textbf{69.4} (\textbf{3.6}) & 0.1611 & 0.4954 & 0.1520 & 0.1807 \\
    \rowcolor{LightBlue} UBD-Unlearning ($T_p = 0.0$) & 58.4 (7.0) & 75.6 (9.8) & 0.2014 & 0.4943 & 0.1641 & 0.2183 \\
    \rowcolor{LightBlue} UBD-Debiasing ($T_p = 0.5$) & 51.6 (13.8) & 70.9 (5.1) & \textbf{0.1490} & \textbf{0.4713} & 0.1299 & \textbf{0.1791} \\
    \rowcolor{LightBlue} UBD-Unlearning  ($T_p = 0.5$) &  58.5 (6.9) & 73.0 (7.2) & 0.2034 & 0.4955 & 0.1639 & 0.2281  \\
    \bottomrule
    \end{tabular}
    }
    \vspace{-0.3cm}
    \label{tab:mathmcqa}
\end{table*}

\section{Experimental Setup}

\subsection{Data}

We investigate data contamination on two MCQ benchmarks as $\mathcal{D}_\text{test}$, namely MMLU-pro \cite{mmlupro} and MATH-MCQA \cite{math_mcqa_2025}. For each benchmark, we split it randomly into two subsets with an equal number of samples, and use one as $\mathcal{D}_\text{dev}$ and another as $\mathcal{D}_\text{eval}$. We construct the rest of $\mathcal{D}_\text{train}$ by collecting another {110k} training samples from Magpie-pro \cite{magpie}, Commonsense-QA \cite{commonsenseqa}, and OpenOrca \cite{OpenOrca} training sets. 
We define the contaminated model and uncontaminated model depending on whether it is trained on $\mathcal{D}_\text{eval}$. The contaminated model is trained on $\mathcal{D}_\text{train} \cup \mathcal{D}_\text{eval}$, and the uncontaminated model is trained on $\mathcal{D}_\text{train} \cup \mathcal{D}_\text{dev}$ to eliminate the gaps in model abilities and domain shifts and only focusing on contamination of the question content itself. 

\subsection{Model and Training}
We evaluate our methods on Llama-3.2-3B-Instruct \cite{llama} and Qwen2.5-3B-Instruct \cite{qwen25}. Training is with LoRA using a rank of 64 and alpha of 128. We perform contamination training using 5 different random seeds corresponding to different initialization and batch ordering, yielding an ensemble of 5 sets of LoRA parameters. To obtain reliable estimates of the uncontaminated model outputs, we also train an ensemble of 5 sets of LoRA parameters for the uncontaminated model and use the averaged output distribution across the ensemble as the reference to measure $D_\text{KL}$ and $D_\text{L1}^*$ for each sample.

\subsection{Baselines}

We primarily compare with approaches that do not require access to a clean model. Specifically, following ITD \cite{itd}, we construct paraphrased questions and choices using GPT-4o. We also adopt a permutation of choices and combine it with paraphrasing as a stronger baseline. Paraphrasing and permutations are performed 3 times, and the probabilities are averaged across the 3 runs to eliminate any potential biasing. These are all widely used dynamic benchmark-style approaches. In addition, we also compare our method to DeconIEP \cite{chai2026benchmarks}, where we use the corresponding original backbone LLMs (i.e. Llama-3.2-3B-Instruct and Qwen2.5-3B-Instruct) to provide reference distributions, and we use the entire benchmark $\mathcal{D}_\text{test}$ as $\mathcal{D}_\text{proxy}$.

\section{Results}

\subsection{Main Results}

We first report the results on MMLU-pro using Llama-3.2-3B-Intrcut and Qwen2.5-3B-Instruct models in Table \ref{tab:llamammlupro}. We report both the overall accuracy differences using residual contamination (RC) \cite{chai2026benchmarks}, and individual sample-level differences using $D_\text{KL}$ and $D_\text{L1}^*$.

\textbf{Gap between dataset-level metrics and sample-level metrics}. The combination of paraphrasing and choice permutation yields the strongest black-box baseline, reducing RC from 17.2 to 8.4 on $D_\text{eval}$ while incurring negligible accuracy difference on $D_\text{dev}$. However, this dataset-level improvement does not translate to the sample level. $D_\text{KL}$ increases by more than 13\% relative to the contaminated model, and $D_\text{L1}^*$ remains largely unchanged. This pattern is consistent across Qwen2.5 results as shown in the second half of Table \ref{tab:llamammlupro}. These findings reveal a fundamental disconnect between dataset-level and sample-level metrics: A reduction in overall performance measures such as RC does not imply that per-sample model behaviour moves closer to that of the uncontaminated model. Consequently, in many cases where accuracy alone is insufficient, explicitly measuring and mitigating sample-level divergence via metrics such as $D_\text{KL}$ and $D_\text{L1}^*$ is essential.

\textbf{UBD achieves consistently better sample-level performance}. While DeconIEP also achieves competitive performance on RC, and better $D_\text{KL}$ and $D_\text{L1}^*$ compared to paraphrasing and permutation methods, the sample-level differences are much larger compared to UBD methods, especially its degradation on $D_\text{dev}$. Meanwhile, when the actual backbone model is unknown, a mismatched clean model may further increase distances. As a result, UBD-Debiasing achieved the best efficacy on $D_\text{eval}$, while UBD-Unlearning achieved the best balance between dev and eval set performance. 

Specifically, with Llama-3.2 model, both UBD-Debiasing and UBD-Unlearning on $D_\text{eval}$ achieved relative reductions of over 40\% $D_\text{KL}$ and $D_\text{L1}^*$ compared to the contaminated model, significantly higher than baseline methods. Meanwhile, UBD-Debiasing maintained a low $D_\text{dev}$ distortion, with $D_\text{KL}$ and $D_\text{L1}^*$ below 0.1. The sample-level distances of the contaminated model is much higher with the Qwen2.5 model in general, and the effect of UBD methods are more obvious. Notably, UBD-Debiasing achieved nearly 60\% relative reductions in $D_\text{KL}$ and around 50\% reduction in $D_\text{L1}^*$.
Results on the MATH-MCQA data are shown in Table \ref{tab:mathmcqa} for both models, with a similar pattern observed. The sensitivity analysis of the hyperparameter $T_p$ is presented in Section \ref{sec:threshold}.

\begin{figure*}[t]
    \centering
    \includegraphics[width=0.9\linewidth]{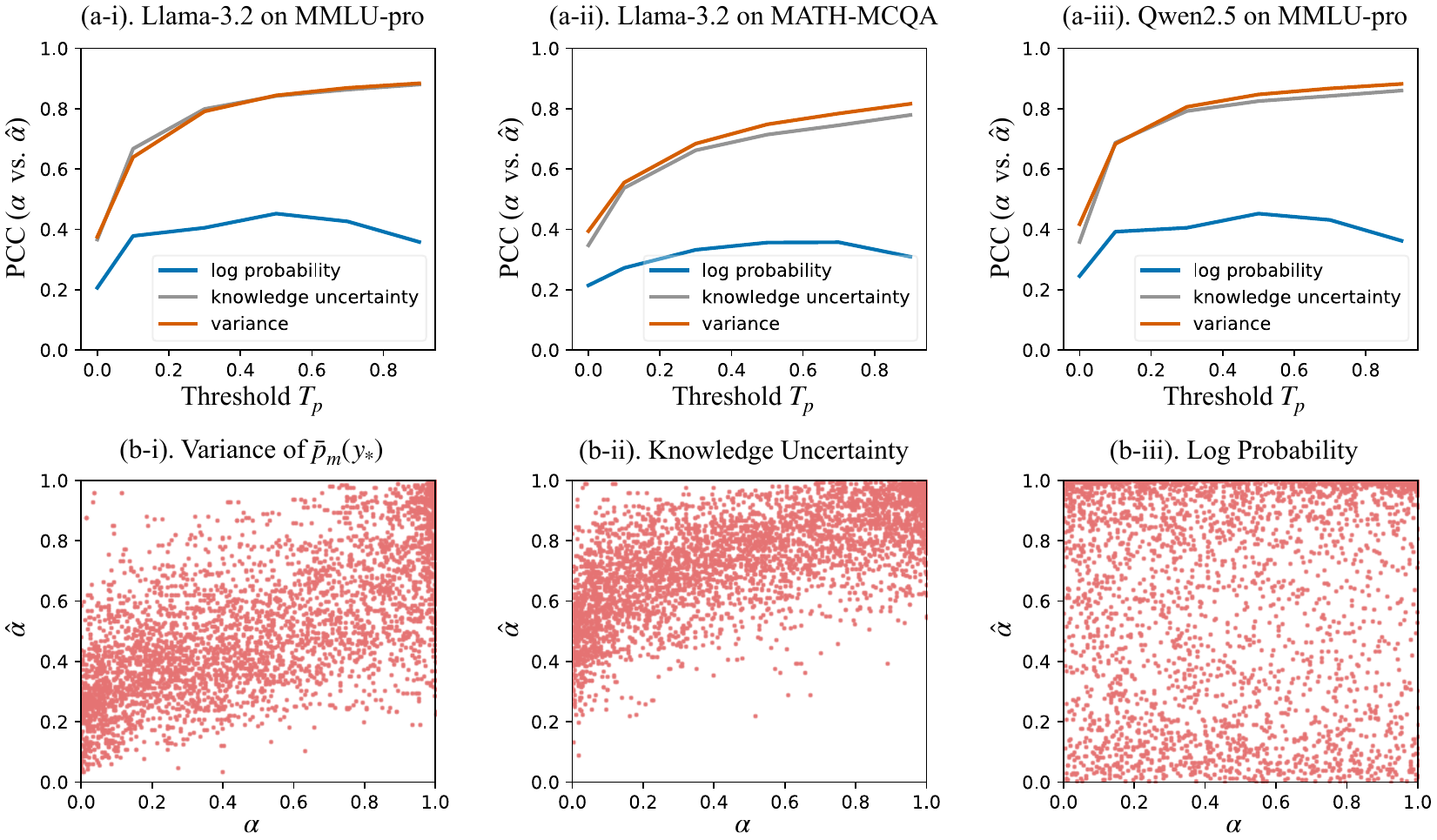}
    \vspace{-0.2cm}
    \caption{Correlation between various contamination indicators against ${\alpha}$. The top 3 plots (a-i to a-iii) show the Pearson Correlation Coefficient (PCC) for data samples against varying thresholds $T_p$. High $\bar{p}^{(i)}(y_*)$ region requires differentiation between hard but memorized samples from easy samples. The lower plots (b-i to b-iii) show the correlation of predicted $\hat{\alpha}$ against oracle $\alpha$ using different indicators on MMLU-pro.}
    \label{fig:alphacorrelation}
\end{figure*}

\subsection{Contamination Level Indicators}
We then compare log probability of the ground-truth option $y_*$ against uncertainty metrics including the standard deviation of $\bar{p}(y_*)$ and knowledge uncertainty $\mathcal{I}(y, \theta)$, as indicators for contamination level. As shown in Fig. \ref{fig:alphacorrelation} (a-i) to (a-iii), uncertainty-based metrics achieved over 0.8 PCC between predicted and ground-truth $\alpha$ when threshold $T_p$ is above $0.5$, whereas log probability stays at a very low correlation level with PCC hardly beyond 0.4. 

When $T_p$ is high, samples contain a mixture of hard but memorized samples ($\alpha\approx 0$) and original easy samples ($\alpha\approx 1$). A good contamination indicator should distinguish between the two types of samples, and apply much larger penalty to the hard but memorized ones. As shown in Fig. \ref{fig:alphacorrelation}, uncertainty-based metrics clearly distinguishes the two with a PCC over 0.9, whereas such difference cannot be derived from log probabilities.

In addition, we provide the scatter plot to show the sample-level correlation between the groundtruth $\alpha$ and $\hat{\alpha}$ predicted with the standard deviation of $\bar{p}_m{(y_*)}$, the knowledge uncertainty, and the log probability of the ground truth option, as shown in Fig. \ref{fig:alphacorrelation} (b-i) to (b-iii) respectively. While both the standard deviation and the knowledge uncertainty achieve similarly high PCC, the standard deviation provides a more linear relationship with a bounded range, making it easy to directly fit with a general linear function in Eqn \eqref{eq:alpha_pred}.

\subsection{Sensitivity to Threshold}
\label{sec:threshold}

\begin{figure}[t]
    \centering
    \hspace{-0.3cm}
    \includegraphics[width=0.98\linewidth]{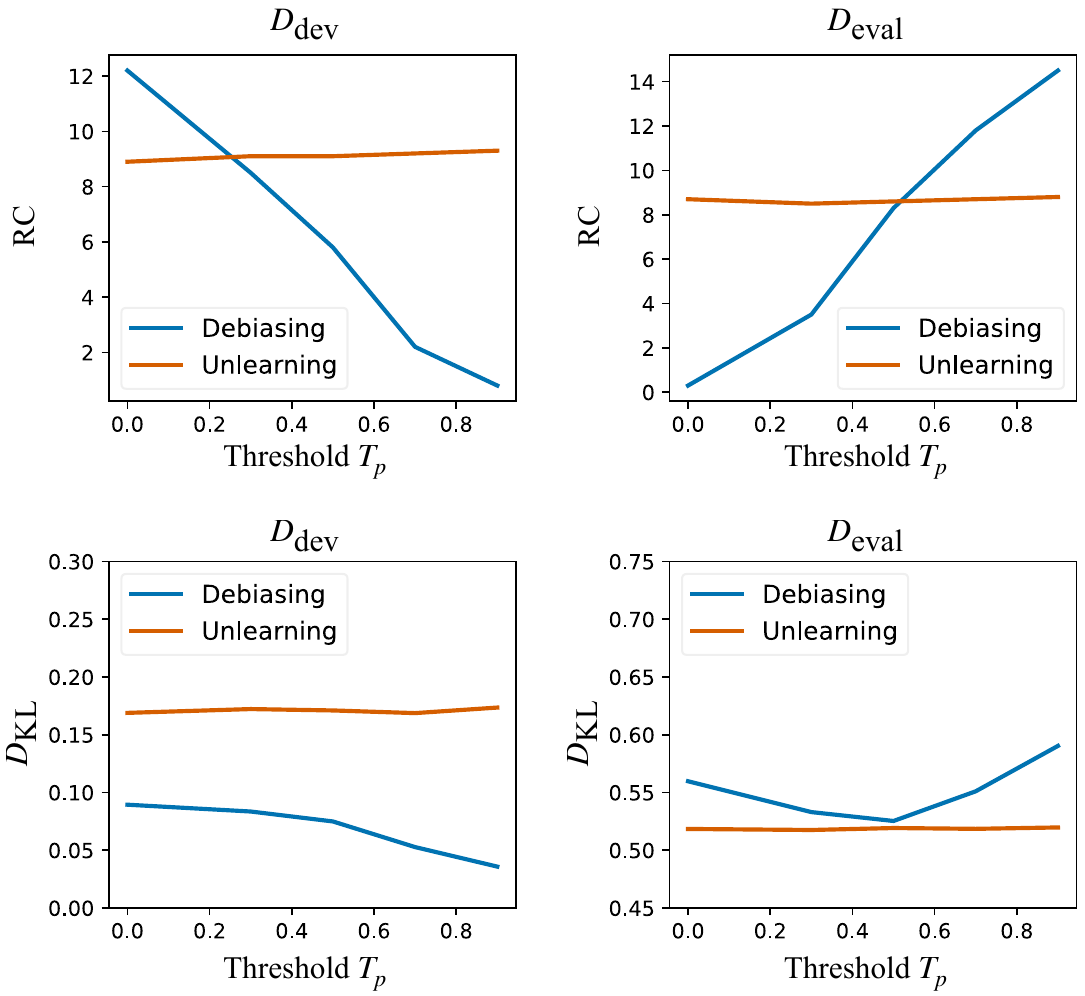}
    \vspace{-0.2cm}
    \caption{Sensitivity of UBD-Debiasing and UBD-Unlearning to threshold $T_p$ based on RC and $D_\text{KL}$.}
    \label{fig:sensitivity}
\end{figure}

We further show the sensitivity of decontamination efficacy against the threshold hyper-parameter using RC and $D_\text{KL}$ in Fig. \ref{fig:sensitivity}. First, while debiasing showed a large sensitivity to the threshold on RC, both methods showed robustness on $D_\text{KL}$ at all threshold levels. Besides, UBD-Unlearing, though performing worse than UBD-Debiasing at the $T_p=0.5$ operating point, shows its performance robustness throughout different thresholds.

\section{Conclusion}

This paper first proposes the sample-level decontamination evaluation, pointing out the gap between overall performance metrics and per-sample differences. $D_\text{KL}$ and $D_\text{L1}^*$ as sample-level performance metrics are proposed. In addition, uncertainty-based decontamination (UBD) using deep ensembles is proposed, which is able to distinguish hard but memorized samples from easy samples. As a result, UBD achieved over 60\% reduction in $D_\text{KL}$ compared to the contaminated model, and outperformed paraphrase and permutation baselines as well as reference-model-based DeconIEP with clear margins.

\section{Limitations}

One of the key limitations is the requirement for ensemble models to estimate the uncertainty. It requires the model providers to release not only the main checkpoint but also an ensemble of checkpoints. In our setting, we fine-tune with LoRA and show that the ensemble of LoRA weights satisfies this requirement, which largely reduces the burden of releasing the number of parameters. Future investigations could explore methods to effectively derive ensembles from a single released model. Another limitation is that UBD-Debiasing in its current form is restricted to classification tasks (MCQ or binary). We believe there exist extensions to be applied to open-ended questions, such as debiasing at each decoding step using the same method, and we leave this for future investigations.


\bibliography{custom}

\appendix

\section{Knowledge Uncertainty}
\label{sec:uncertainty}
Following \cite{kendall2017uncertaintiesneedbayesiandeep,malinin,malinin2}, we quantify knowledge uncertainty via the mutual information between the model output and the ensemble parameters, approximated as:
\begin{align}
\small \hspace{-0.3cm}
    \mathcal{I}(y, \theta \mid x^{(i)}) & \approx \mathcal{H}\!\left[\frac{1}{M}\sum_{m=1}^{M} P(y \mid x^{(i)}, \bar{\theta}_m)\right] \nonumber \\
    &- \frac{1}{M}\sum_{m=1}^{M} \mathcal{H}\!\left[P(y \mid x^{(i)}, \bar{\theta}_m)\right],
\end{align}
where $\mathcal{H}[\cdot]$ denotes the entropy of a distribution and $\mathcal{I}[\cdot]$ is the mutual information. Knowledge uncertainty measures the degree of disagreement among ensemble members. Importantly, uncertainty from ensembles distinguishes the hard but memorized sample from easy samples. When the probability of the ground-truth is high, hard but memorized sample exhibits high inconsistency across ensembles, yielding higher uncertainty than easy samples. 



\end{document}